\documentclass[final,5p,times,twocolumn]{elsarticle}

\journal{Physics Letters B}

\usepackage{graphicx}
\usepackage{animate}
\usepackage{float}
\usepackage{ulem}
\usepackage{comment}
\usepackage[T1,T2A]{fontenc}
\usepackage[utf8]{inputenc}
\usepackage[english]{babel}
\usepackage{amsmath}
\usepackage{amssymb}
\usepackage{lineno}
\usepackage{footmisc}
\usepackage{mathtools}
\usepackage{perpage}
\MakePerPage{footnote}
\usepackage{tikz}
\usepackage{multirow}
\usepackage{bm}
\usepackage{dcolumn}
\newcommand{\ys}[1]{{\textcolor{blue}{{#1}}}}

\newcommand{\bea}{\begin{eqnarray}}
\newcommand{\eea}{\end{eqnarray}}
\newcommand{\be}{\begin{equation}}
\newcommand{\ee}{\end{equation}}

\usepackage{pgfplots}
\pgfplotsset{compat=1.15}
\usepackage{mathrsfs}
\usetikzlibrary{shapes.geometric, arrows}

\usepackage{xifthen}
\newcommand*\diff{\mathrm{d}} 
\newcommand*\ldiff[2][]{ \ifthenelse{\isempty{#1}}{ \diff #2}{\diff^#1#2} \,} 
\let\limitint\int 
\renewcommand{\int}{\limitint \!} 

\allowdisplaybreaks

\tikzstyle{startstop} = [rectangle, rounded corners, minimum width=1.8cm, minimum height=1cm,text centered, draw=black, fill=red!30]

\tikzstyle{arrow} = [thin,->,>=stealth]

\addto\captionsenglish{}

\renewcommand{\L}{\mathcal{L}}

\begin{document}
\begin{frontmatter}

\title{Localized Electromagnetic Perturbations on Vector Solitons}

\author[first]{Yulia Galushkina}
\affiliation[first]{organization={Institute for Nuclear Research of RAS},
            addressline={prospekt 60-letiya Oktyabrya 7a}, 
            city={Moscow},
            postcode={117312}, 
            country={Russia}}

\author[first,second]{Eduard Kim}
\affiliation[second]{organization={Moscow Institute of Physics and Technology},
            addressline={Institutsky lane 9}, 
            city={Dolgoprudny},
            postcode={141700}, 
            state={Moscow region},
            country={Russia}}
 
\author[first]{Emin Nugaev}

\author[fourth,fifth]{Yakov Shnir}
\affiliation[fourth]{organization={BLTP, JINR},
            addressline={6 Joliot-Curie St}, 
            city={Dubna},
            postcode={141980}, 
            state={Moscow region},
            country={Russia}}
\affiliation[fifth]{organization={Institute of Physics, Carl von Ossietzky University Oldenburg},
            addressline={Carl-von-Ossietzky-Straße 11}, 
            city={Oldenburg},
            postcode={26129}, 
            state={Lower Saxony},
            country={Germany}}

\begin{abstract}
   Using effective field theory approach one can describe localization of electromagnetic field on a non-topological soliton. Pursuing this aim we consider the $U(1)$ gauge theory with gauge kinetic coupling to a
   self-interacting complex neutral Proca field. The model possesses single energy scale given by the vector boson mass. Considering spherically symmetric stationary configurations, we study vibrational modes of the Proca field on the background of the soliton and discuss their  properties.
\end{abstract}


\end{frontmatter}

\section{Introduction }
 \label{Introduction}

There has been significant progress in recent decades in the
analysis of spatially localized regular solutions of nonlinear equations, the solitons,
for a review, see e.g. \cite{manton2004topological,shnir2018topological}.
The solitons exist in various space-time dimensions, they represent static particle-like objects or  miscellaneous
traveling waves. Such configurations appear in various contexts, for example, they play important roles in condensed
matter systems \cite{bishop1980solitons}, classical and quantum field theory \cite{weinberg2012classical}, supersymmetric theories
\cite{shifman2009supersymmetric} and cosmology \cite{kibble1980some,vilenkin1994cosmic}. Most generally, in classical field theory
the solitons fall into
two distinct classes: (i) topological and (ii) non-topological solitons. While topological solitons are classified according to
the value of the topological charge that, in most cases originates from the spontaneous symmetry breaking, non-topological
solitons carry a conserved Noether charge which is related to a symmetry of the corresponding Lagrangian. A paradigmatic
example are Q-balls, they are localized configurations of a self-interacting complex scalar field with a stationary oscillating
phase \cite{Rosen:1968mfz,Friedberg:1976me,Coleman:1985ki,Radu:2008pp}.

A suggestive idea in field theory is the possibility of making a soliton that supports a localized electromagnetic field. 
However, there is a number of obstacles to the localization of gauge fields that have been encountered in studies in the approach with extra spatial dimensions (see  \cite{Rubakov:2001kp} for review and \cite{Dvali:2000rx} for details).

Contrary to the massless electromagnetic field, 
complex Proca field with the global Abelian symmetry \cite{Loginov:2015rya}
can form non-topological solitons due to the self-interaction potential\footnote{Ambiguities related to singularities of Proca field were considered in \cite{Zhang:2021xxa,Clough:2022ygm}. To avoid these issues, we study solitons in a framework of low-energy effective field theory (EFT).}. Moreover, it was pointed out that massive vector particles and compact objects formed by a Proca field may be considered as candidates for dark matter \cite{Gorghetto_2022}. A usual portal coupling the Maxwell electrodynamics to a hidden Proca sector is the gauge kinetic coupling \cite{Okun:1982xi,Holdom:1985ag}. This mechanism provides an opportunity to localize electromagnetic field on a vector soliton, as suggested recently \cite{Galushkina:2024knf}.
In this model, electromagnetic field localization is obtained on Proca balls, non-topological solitons with a harmonic time-dependence of the massive complex vector field \cite{Loginov:2015rya}. Peculiar feature of Proca balls is that they may exist for a very small
range of values of the angular frequencies\footnote{Coupling to gravity may
stabilize the soliton, this range becomes extended \cite{Brito:2015pxa}.}.

An interesting aspect of non-topological solitons is related with their perturbations, typically, apart some set of zero modes, a few
localized vibrational modes may appear in the spectrum of second-order small fluctuation operator
\cite{1970JMP....11.1336A, Smolyakov:2017axd,Kovtun:2018jae, Dorey:2021mdh}. 
Quite importantly, quasinormal modes
or resonant excitations (for a review see e.g. \cite{Konoplya:2011qq,Ching:1998mxl}),
may also occur in the spectrum of excitations of a soliton\footnote{A vast literature is devoted to
the quasinormal modes of self-gravitating solitons and hairy black holes,which we will not discuss here.}
\cite{Forgacs:2003yh,Bizon:2007qz}.
Excitations of the internal and/or quasinormal modes in various processes of interactions of solitons lead to
many surprising results, like inelastic collisional dynamics with a chaotic structure and formation of bound states
\cite{Campbell:1983xu,Anninos:1991un}, boundary collisions \cite{Dorey:2015sha}, 
the effect of negative radiation pressure \cite{Romanczukiewicz:2003tn,Forgacs:2008az}, appearance of spectral walls \cite{Adam:2019xuc}
and resonance scattering through the quasinormal modes \cite{Dorey:2017dsn}. 
Linear perturbations were studied both for Proca balls/stars \cite{Herdeiro:2023wqf,Santos:2024vdm}
and gauged solitons \cite{Alonso-Izquierdo:2024bzy,Kinach:2024qzc,Panin:2016ooo}.

The aim of this Letter is to analyze the spectrum of linearized perturbations of non-topological solitons
in a theory of complex neutral self-interacting Proca field with the global $U_g(1)$ symmetry coupled to electromagnetic field.
In Sec.\ref{Sec.2}, we discuss the model and vector solitons therein. Sec.\ref{Sec.3} provides analysis of spherically symmetric perturbations of solitons. We demonstrate localized mode of electromagnetic field and describe spectrum of perturbations within the limits of applicability of EFT.

\section{Background}\label{Sec.2}
The presence of matter fields may significantly affect  properties of electromagnetic waves.
Celebrated examples are the scalar electrodynamics and the Abelian Higgs model. The scalar electrodynamics provides a simplest possible example of a renormalizable and gauge-invariant theory. However, in EFT approach one can construct an interaction that only has quadratic form in $F_{\mu \nu}$, so that the gauge potential $A_\mu$ does not appear explicitly in the field equations. Thus, electromagnetic field can be classically integrated out using equations of motion.

Let us consider a (3+1)-dimensional theory of the complex massive neutral field $V^\mu$ coupled to the electromagnetic field tensor $F_{\mu\nu}$
\begin{equation}
\label{L}
\mathcal{L} = -\frac{1}{4} F_{\mu\nu}F^{\mu\nu} -\frac{1}{2} V^*_{\mu\nu}V^{\mu\nu} + \frac{ i \gamma}{2}F_{\mu\nu}W^{\mu\nu} - U(V^\nu, V^{*\mu}) \;, 
\end{equation}
where $F_{\mu\nu} = \partial_\mu A_\nu - \partial_\nu A_\mu$, $V_{\mu\nu} = \partial_\mu V_\nu - \partial_\nu V_\mu$, and purely imaginary combination $W_{\mu\nu} =  V^*_\mu V_\nu - V^*_\nu V_\mu$. Coupling $\gamma$ is a dimensionless constant, and \(U(V^\nu, V^{*\mu})\) is a $U(1)$-invariant potential of self-interaction. Here we require the global $U(1)$ symmetry of this theory and the gauge-invariant coupling of $V^\mu$ to $F_{\mu\nu}$ along with $P$-parity. The third term in~(\ref{L}) provides magnetic moment of a vector boson, \(\gamma/2M\) (see \cite{Pauli:1941zz}). Spherically symmetric solitons in this model (with 4th-order self-interaction of vectors) were studied in the paper \cite{Galushkina:2024knf}.

A legitimate approach to model (\ref{L}) is to consider it as an effective theory describing an interaction of the electromagnetic field with certain (composite) bosons of non-zero spin. Particularly, experiments with condensate of dipolar bosonic gases \cite{pargellis1991monopole, Lahaye:2009qqr} may allow observation of electromagnetic field localization. For this reason we will consider self-interaction in (\ref{L}) which is consistent in EFT approach in the non-relativistic (NR) regime.

As mentioned above, the key property of the model (\ref{L}) is that at the classical level the electromagnetic field can be integrated out completely. Indeed, in such a case the Maxwell equations take the form
\begin{equation}
\label{E_to_W}
\partial_\mu F^{\mu\nu} = i \gamma \partial_\mu W^{\mu\nu} \;
\end{equation}
and, for configurations with trivial boundary conditions at spatial infinity, we obtain:
\begin{equation}
\label{E_to_W2}
F^{\mu\nu} = i \gamma W^{\mu\nu} \;.
\end{equation}
Classically, this relation significantly simplifies equations of motion of the model (\ref{L}). Certainly, situation is different and more complicated in a consistent QFT, as the gauge field $A_{\mu}$ should be quantized as a truly dynamical field together with other degrees of freedom.

After discussing the model with arbitrary self-interaction, we set \(U(V^\nu, V^{*\mu})\) to be the most general P-even 4th order potential:
\begin{equation}
\label{U1}
U = - M^2V^*_{\mu}V^{\mu} -\frac{\alpha}{2}(V^*_{\mu}V^{\mu})^2 - \frac{\beta}{2}(V^*_{\mu}V^{*\mu})(V_{\nu}V^{\nu})
\end{equation}
with \(\alpha\), \(\beta\) dimensionless constants. 
This theory has consistent description in the zero limit of all dimensionless coupling constants and the only scale parameter is the mass of the free bosons $M$. This feature of (\ref{U1}) significantly simplifies the analysis of the applicability of the EFT. Details of UV-completion, i. e. presence of higher order terms of interaction or additional fields (e. g. like in \cite{Loginov:2015rya, Herdeiro:2023lze}), do not change our results for NR solitons in the limit of small higher order EFT interactions.

Substituting (\ref{E_to_W}) into (\ref{L}), we arrive at the reduced Lagrangian containing the vector field only,
\begin{equation}
\begin{aligned}
\label{eff_model}
& \L =  -\frac{1}{2} V^*_{\mu\nu}V^{\mu\nu} - \tilde{U}(V^\nu, V^{*\mu}) \;, \\
& \tilde{U} = -  M^2V^*_{\mu}V^{\mu}  - \frac{\tilde{\alpha}}{2} (V^*_{\mu}V^{\mu})^2 - \frac{\tilde{\beta}}{2}(V^*_{\mu}V^{*\mu})(V_{\nu}V^{\nu}) \;, \\  
\end{aligned}
\end{equation}
where  \( \tilde{\alpha} = \alpha - \gamma^2\) and \( \tilde{\beta} = \beta + \gamma^2\). In this paper we restrict ourselves to \(\tilde{\alpha} > 0\) (see \cite{Galushkina:2024knf} for the discussion about coupling constants). Contrary to the scalar case \cite{1970JMP....11.1336A}, one can obtain stable solitons in a vector theory with only quartic couplings. 

It is convenient to rescale the model by introducing
the dimensionless units as follows,
\begin{equation}
\label{var}
V^\nu = \frac{M\Tilde{V}^\nu}{\sqrt{\Tilde{\alpha}}} \;, \,\,\, \kappa = \frac{\Tilde{\beta}}{\Tilde{\alpha}} \;, \,\,\, x_ix^i = \frac{r^2}{M^2} \;, \,\,\, t = \frac{\tau}{M} \;.
\end{equation}
The action $S$ of the theory becomes $S=\Tilde{\alpha}^{-1}\tilde{S}$ where $\Tilde{S}$ depends only on $\kappa$. We require $\Tilde{\alpha}/4\pi\ll 1$ for the validity of semiclassical approach.

After replacing variables the field equations become
\begin{equation}
\label{vectorEOM}
\partial_\mu \Tilde{V}^{\mu\nu} + \Tilde{V}^\nu + (\Tilde{V}^*_\mu \Tilde{V}^\mu) \Tilde{V}^\nu + \kappa (\Tilde{V}_\mu\Tilde{V}^\mu) \Tilde{V}^{*\nu} = 0 \;,
\end{equation} 
There is also a constraint:
\begin{equation}
\label{Constraint}
\partial_\nu (\Tilde{V}^\nu + (\Tilde{V}^*_\mu \Tilde{V}^\mu) \Tilde{V}^\nu + \kappa (\Tilde{V}_\mu\Tilde{V}^\mu) \Tilde{V}^{*\nu}) = 0 \;,
\end{equation} 
Note that the only parameter in these equations is \(\kappa\), which means that theories with different \(\Tilde{\alpha}\) and the same \(\kappa\) are classically equivalent.

To obtain spherically symmetric hedgehog-like solitons we consider the following radial ansatz studied in many papers (e. g. \cite{Loginov:2015rya, Herdeiro:2023lze, Galushkina:2024knf}): 
\begin{equation}
\label{ansatz}
\Tilde{V}_{0} = i u(r) e^{-i \mathrm{w} \tau} \,\, , \,\,
\Tilde{V}^{i} = \frac{\Tilde{x}^{i}}{r} v(r) e^{-i \mathrm{w} \tau},
\end{equation}
where $\Tilde{x}^i = Mx^i$, \(u(r)\), \(v(r)\) are profile functions of the vector field, and \(\mathrm{w}\) is the angular frequency in units of the field mass. For bound states the inequality \(\mathrm{w} < 1\) is fullfilled.
In terms of this ansatz, (\ref{vectorEOM}) becomes
\begin{equation}
\label{ans_eq}
    \begin{aligned}
&u'' + \frac{2}{r} u' - \mathrm{w}(v' + \frac{2}{r} v) \\
& \qquad\qquad - u - (u^2 - v^2)u - \kappa (u^2 + v^2) u = 0 \;, \\
& \mathrm{w} u' + (1 - \mathrm{w}^2) v \\
& \qquad\qquad + (u^2 - v^2)v - \kappa (u^2 + v^2) v = 0 \;,
    \end{aligned}
\end{equation}
where prime means the derivative with respect to $r$. The constraint (\ref{ans_eq}) combined with the second equation from (\ref{Constraint}) takes the form:
\begin{equation}
\begin{split}
\label{denominator}
v' & = \frac{\mathrm{w}}{1 + (u^2 - v^2) - \kappa (u^2 + v^2) - 2v^2(1+\kappa)} \\ 
& \times \biggl(- \frac{2}{\mathrm{w}}(1 -\kappa)uu'v + \frac{2(u' - \mathrm{w}v)}{r}  \\ 
& -u - u (u^2 - v^2) -\kappa u (u^2 + v^2) \biggr) \;.
\end{split}
\end{equation}
If the denominator in (\ref{denominator}) vanishes, the equations of motion become inconsistent. This property of vector theories was studied in \cite{Mou:2022hqb, Clough:2022ygm}. However, in the non-relativistic limit the problem with denominator does not occur because \(u^2\) and \(v^2\) are small comparing to 1. Moreover, for solitons in the model we consider it was shown in \cite{Galushkina:2024knf} that for $-1<\kappa\lesssim 0.6$ the denominator never vanishes and we obtain full branches of solitons with thin-wall regime at $\mathrm{w} \to \mathrm{w}_{min}$. For solitons with $-0.6\lesssim\kappa < 0$, smoothness of solution is lost at $\mathrm{w} \to \mathrm{w}_{min}$ and there are no thin-wall solitons. For \(\kappa > 0\) all the solitons are unstable.

The global $U(1)_g$ Noether charge and energy of a soliton, found in (\cite{Galushkina:2024knf}), are of the form
\begin{align}
\label{QQ}
    & Q = \frac{8\pi}{\Tilde{\alpha}} \limitint_0^\infty v(\mathrm{w} v - u' )r^2 \: \diff r \;, \\ 
    & E = \frac{4\pi M}{\Tilde{\alpha}} \limitint_0^\infty  \bigl( (\mathrm{w} v - u')^2 \nonumber + \\ 
\label{EE}
    & \quad + \bigl( u^2 + v^2 \bigr) + \frac{1}{2}(u^2 - v^2)(3 u^2 + v^2)  \\
    & \quad + \frac{\kappa}{2}(u^2 + v^2)(3u^2 - v^2)\bigr)r^2 \: \diff r \;. \nonumber
\end{align}

We should stress that r.h.s of Eq.(\ref{E_to_W}) provides additional conserving current
\begin{equation}
j_{T}^\nu = i \partial_\mu W^{\mu\nu} \;.
\end{equation}
It is worth noticing that the conservation law for $j_{T}$ is of topological nature, since $W^{\mu \nu}$ is antisymmetric and does not rely on equations of motion.
Current $j_{T}$ can be treated as a source of electromagnetic field. Our choice of coupling of vector field to $F^{\mu \nu}$ ensures the absence of total electric or magnetic charges for solitons.

Numerical solutions of the equations (\ref{vectorEOM}) were discussed in \cite{Galushkina:2024knf}. The obtained solitons demonstrate the electromagnetic field localization due to equation (\ref{E_to_W2}). Fields profiles \(V^{\mu}\) decrease exponentially at spatial infinity and so does the electric field. This is the property of the model which provides electromagnetic field localization. In this paper we will show that the electromagnetic field localization is also observed for perturbation modes.

\begin{figure}[t]
\includegraphics[width=0.9 \linewidth, height = 6 cm]{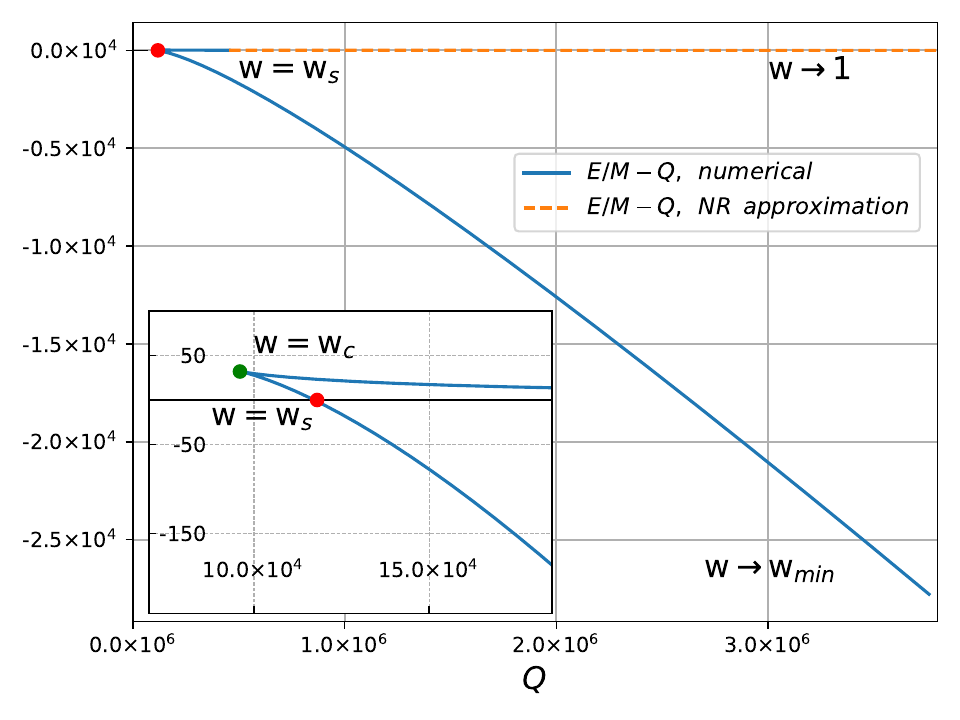}
\caption{Integral characteristic $(E/M - Q)$ plotted versus $Q$ for vector solitons (\ref{ansatz}) in the theory (\ref{eff_model}) with $\tilde{\alpha} = 1$ and $\kappa = \tilde{\beta}/\tilde{\alpha}= -0.9$. Marked values of \(\mathrm{w}\): \(\mathrm{w}_{min} \approx 0.99\), \(\mathrm{w}_{s} = 0.99806\), \(\mathrm{w}_{c} = 0.99918\). The upper branch of solitons is partially calculated on the obtained numerical solutions and partially with the formulae of non-relativistic scaling of E and Q (see \cite{Loginov:2015rya}). }
\label{E - MQ}
\end{figure}

In the present paper we fix \(\kappa = - 0.9\) and study oscillating perturbations for solitons with different \(\mathrm{w}\). We obtain soliton profiles with 4th order Runge-Kutta method, controlling the precision with the analysis of step size and box size dependence of results and also using the equality \(
\partial E/{\partial\mathrm{w}} = \mathrm{w}\partial Q/\partial \mathrm{w}\). Results for E, Q and critical points (\(\mathrm{w}_c\), \(\mathrm{w}_s\) on Fig.\ref{E - MQ}) are accurate to the 5th significant digit.

For \(\kappa = - 0.9\), the thin wall regime of solitons is obtained near $\mathrm{w}_{min} \approx 0.99 $. The region of linear and kinematic stability of solitons is determined by their \(E(\mathrm{w})\) and \(Q(\mathrm{w})\) dependency. The criterion of kinematic stability is fulfilled if $(E - MQ) < 0$ which means that the energy of the soliton is less than the energy of free particles with the same Q. The function \((E - MQ)(Q)\) is shown in Fig.\ref{E - MQ}. In the region \(\mathrm{w} < \mathrm{w_s}\), where \(\mathrm{w_s} = 0.99806\) is the point at which \(E - MQ = 0\), the solitons are kinematically stable. The condition of linear stability is given by Vakhitov-Kolokolov criterion (\(dQ / d \mathrm{w} < 0\)). Indeed, using the same arguments as in the scalar case \cite{Kovtun:2018jae} one can find decay modes for the upper branch of solutions (\(\mathrm{w} > \mathrm{w_c}\) in Fig. \ref{E - MQ}, \(\mathrm{w}_c = 0.99918\)). The similar results for boson stars were numerically obtained in \cite{Santos:2024vdm}.


\section{Spherically symmetric perturbations}\label{Sec.3}

After discussing the background, we move on to studying the linear perturbations of the solitons. In terms of background field and perturbations the fields can be written as:
\begin{equation}
\label{EM_perturb}
\Tilde{F}^{\mu\nu} = \Tilde{F}_b^{\mu\nu} + \Tilde{F}_p^{\mu\nu}, \,\,\ \Tilde{V}^\mu = \Tilde{V}_b^\mu + \Tilde{V}_p^\mu
\end{equation}
where \(\Tilde{F}^{\mu\nu} = (\sqrt{\Tilde{\alpha}}/M^2) F^{\mu\nu}\); \(\Tilde{F}_b^{\mu\nu}\), \(\Tilde{V}_b^\mu\) are background fields and \(\Tilde{F}_p^{\mu\nu}\), \(\Tilde{V}_p^\mu\) are perturbations. 

First, we use the fact that the equality (\ref{E_to_W2}) is fulfilled: both background and perturbation fields have trivial boundary conditions at spatial infinity. In the linear order it takes the form: 
\begin{equation}
\label{E_to_W_1order}
\Tilde{F}_p^{\mu\nu} =  \frac{\rm{i}\gamma}{\sqrt{\Tilde{\alpha}}} (\Tilde{V}_b^{*\mu} \Tilde{V}_p^\nu - \Tilde{V}_p^{*\nu} \Tilde{V}_p^\mu) +  \frac{\rm{i} \gamma}{\sqrt{\Tilde{\alpha}}} (\Tilde{V}_p^{*\mu} \Tilde{V}_b^\nu - \Tilde{V}_p^{*\nu} \Tilde{V}_b^\mu) \;.
\end{equation}

\(\Tilde{F}_p^{\mu\nu}\) can be integrated out, so we can focus on studying perturbations in the effective theory of self-interacting vector field, similarly to the approach applied to background fields. As it is seen in (\ref{E_to_W_1order}), electromagnetic field perturbations are determined by the linear perturbation of \(\Tilde{V}^{\mu}\).

Let us consider following spherically symmetric ansatz for oscillating perturbation modes:
\begin{multline}
\label{spherical_ansatz1}
\Tilde{V}^0_p = i e^{-i\mathrm{w}\tau}(\chi_1(r) e^{-i\lambda\tau} + \chi_2(r) e^{i\lambda\tau}), \\ \Tilde{V}^i_p = \frac{\Tilde{x}^i}{r}e^{-i\mathrm{w}\tau}(\phi_1(r) e^{-i\lambda\tau} + \phi_2(r) e^{i\lambda\tau}),
\end{multline}
where \(\lambda\) is a real frequency parameter of oscillations, and \(\chi_1\), \(\chi_2\), \(\phi_1\) and
\(\phi_1\) are real profile functions of oscillating modes. This ansatz is similar to one usually used for scalar Q-ball perturbations (see \cite{1970JMP....11.1336A,Smolyakov:2017axd,Kovtun:2018jae}). For vector theory (Proca stars) it was studied in \cite{Santos:2024vdm}.

The equations for linear perturbations take the form:
\begin{equation}
\label{ans_eq_perturb_1}
    \begin{aligned}
&\chi_1'' + \frac{2}{r} \chi_1' - (\mathrm{w} + \lambda)(\phi_1' + \frac{2}{r} \phi_1) - \chi_1 -  \\
& \qquad  - [(2u^2 - v^2)\chi_1 - uv\phi_1 + u^2\chi_2 - uv\phi_2] - \\
& \qquad\qquad - \kappa [(u^2 + v^2)\chi_2 + 2(u^2\chi_1 + uv\phi_1)] = 0 \;, \\\\
&\chi_2'' + \frac{2}{r} \chi_2' - (\mathrm{w} - \lambda)(\phi_2' + \frac{2}{r} \phi_2) - \chi_2 -  \\
& \qquad  - [(2u^2 - v^2)\chi_2 - uv\phi_2 + u^2\chi_1 - uv\phi_1] - \\
& \qquad\qquad - \kappa [(u^2 + v^2)\chi_1 + 2(u^2\chi_2 + uv\phi_2)] = 0  \;, 
    \end{aligned}
\end{equation}
\begin{equation}
\label{ans_eq_perturb_2}
    \begin{aligned}
& (\mathrm{w} + \lambda) \chi_1' + (1 - (\mathrm{w} + \lambda)^2)\phi_1 + \\
& \qquad + [(u^2 - 2v^2) \phi_1 + uv\chi_1 + uv\chi_2 - v^2 \phi_2] - \\
& \qquad\qquad - \kappa [(u^2 + v^2) \phi_2 + 2(uv\chi_1 + v^2\phi_1 )] = 0 \;, \\\\
& (\mathrm{w} - \lambda) \chi_2' + (1 - (\mathrm{w} - \lambda)^2)\phi_2 + \\
& \qquad + [(u^2 - 2v^2) \phi_2 + uv\chi_2 + uv\chi_1 - v^2\phi_1] - \\
& \qquad\qquad - \kappa [(u^2 + v^2) \phi_1 + 2(uv\chi_2 + v^2 \phi_2 )] = 0 \;. \\\\
    \end{aligned}
\end{equation}
From the constraint (\ref{Constraint}) we get:
\begin{equation}
\label{constr_perturb}
    \begin{aligned}
& (\mathrm{w}+ \lambda)A_1 + \frac{2}{r} B_1 + \phi_1' C_1 + \phi_2' C_2 + D_1 = 0\; \\
& (\mathrm{w} - \lambda)A_2 + \frac{2}{r} B_2 + \phi_1' C_2 + \phi_2' C_1 + D_2 = 0\; ,
    \end{aligned}
\end{equation}
where we introduced notations $A_{1},A_{2},...,D_{2}$ for convenience. Their explicit forms are given in \ref{app2}. 

\begin{figure}[h]
\includegraphics[width=0.85 \linewidth, height = 8 cm]{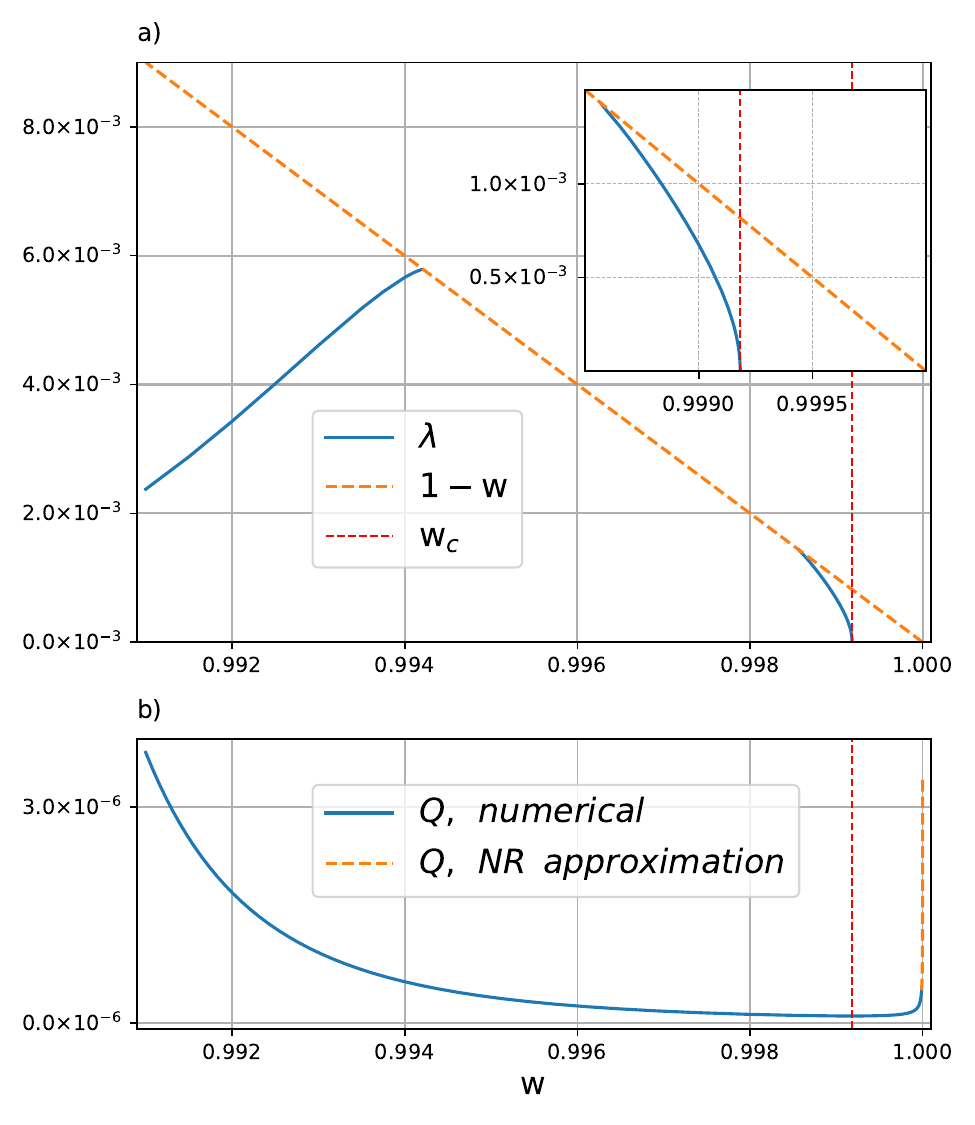}
\caption{ a) Oscillation parameter $ \lambda$ as a function of $\mathrm{w}$ for spherically symmetric perturbations (\ref{spherical_ansatz1}) in the theory (\ref{eff_model}) with $\tilde{\alpha} = 1$ and $\kappa = \tilde{\beta}/\tilde{\alpha}= -0.9$. The point of the cusp, where \(dQ/d\mathrm{w} = 0\), is at \(\mathrm{w}_c = 0.99918\). b) Q as a function of $\mathrm{w}$, with a minimum at \(\mathrm{w}_c\).}
\label{lambda_w}
\end{figure}

We solve these equations numerically with the methods, described in \ref{app1}, to obtain localized perturbations (exponentially decreasing at spatial infinity) and corresponding values of \(\lambda\). The results for \(\lambda(\mathrm{w})\)\footnote{Results for \(\lambda\) are accurate to the 4th significant digit, see \ref{app1}.} are shown in Fig.\ref{lambda_w}. Starting from the vicinity of the cusp (\(\mathrm{w} \rightarrow \mathrm{w_c}\), where \(dQ/d\mathrm{w} \rightarrow 0\) see Fig.\ref{E - MQ}, Fig.\ref{lambda_w}), we found one localized oscillating mode. The form of this mode follows from the expansion in \(\lambda\) near cusp, similarly as in scalar case (see \cite{Kovtun:2018jae, Smolyakov:2017axd}), and this fact was used to validate our numerical results. This result is similar to the case of Proca stars, see \cite{Santos:2024vdm}. However, at $\mathrm{w}=0.9986$ the size of the mode rapidly increases and localized state disappears. 
Indeed, one can notice that delocalization occurs when $\lambda(w)$ hits the border of $(1-w)$, where the condition of exponential asymptotic behavior for oscillating modes is violated.

\begin{figure}[h]
\includegraphics[width=0.8 \linewidth, height = 11 cm]{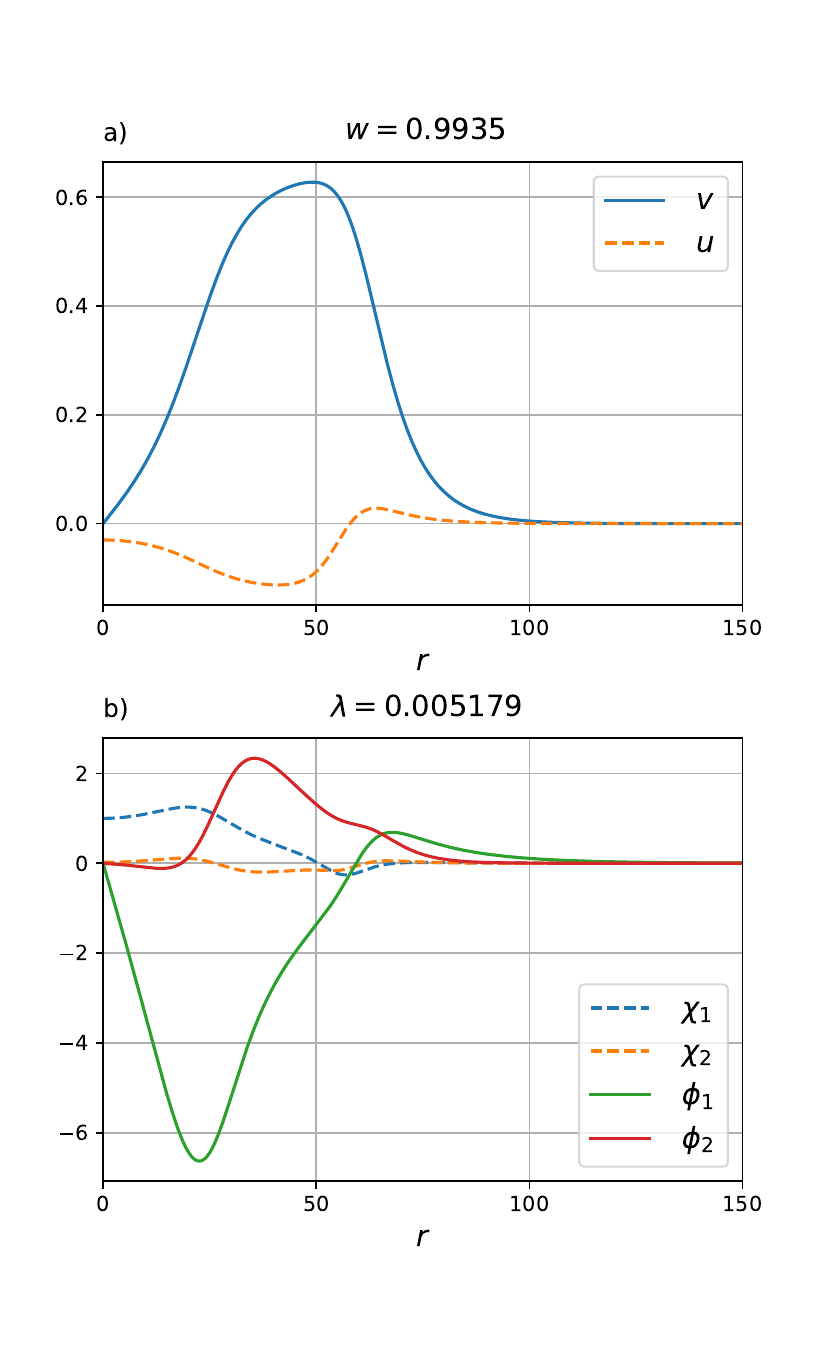}
\caption{a) Background vector field profile (\ref{ansatz}) at \(\kappa = - 0.9\) and \(\mathrm{w} =  0.9935\). b) Oscillating mode profile (\ref{spherical_ansatz1}) at the same parameters, \(\lambda = 0.005179\).}
\label{mode}
\end{figure}

At $\mathrm{w}=0.9942$ localized state revives back into the discrete spectrum, see the left side of Fig.\ref{lambda_w}. In the contrast to the modes near the cusp, these localized states lie in the kinematically stable region (\(E < MQ\)). Profile of the perturbation mode at \(\mathrm{w} = 0.9935\) is shown in Fig.\ref{mode}.

Similar spectrum in scalar theory was studied analytically in \cite{Kovtun:2018jae} with more than one localized states. However, in our theory may be more localized states in the thin-wall limit (w < 0.99), numerical analysis of this study is highly imprecise in this region.

\begin{figure}[h]
\includegraphics[width=0.8 \linewidth, height = 11 cm]{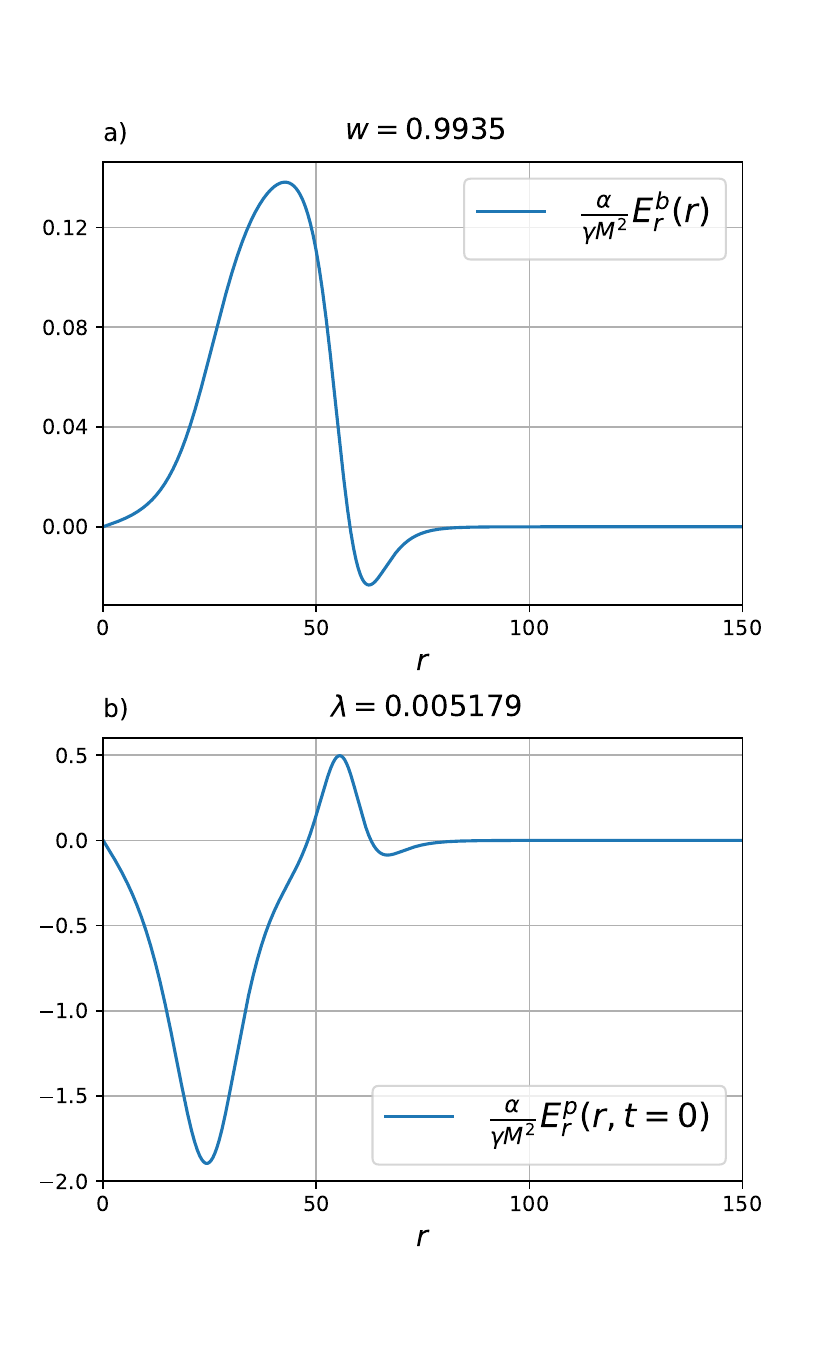}
\caption{a) Background radial electric field profile (\ref{E_field_back}) at \(\kappa = - 0.9\) and \(\mathrm{w} =  0.9935\). b) Oscillating mode radial electric field profile (\ref{E_field_pert}) at the same parameters, \(\lambda = 0.005179\).}
\label{mode_electr}
\end{figure} 

The obtained oscillating modes of vector solitons provide the electric field localization, according to (\ref{E_to_W_1order}). The magnetic field is equal to zero\footnote{If we do not restrict ourselves to P-even interactions, we can modify the discussed model to obtain magnetic field localization (see \cite{Galushkina:2024knf}) and corresponding perturbation modes. In this case, radial magnetic field has similar profiles as shown in Fig.\ref{mode_electr} for electric field.} on these solitons and their perturbations. The electric field of a spherically symmetric soliton with spherically symmetric perturbations is
\begin{equation}
\label{E_field}
\vec{E}(x^i, t) = \vec{n} E_r, \,\,\ E_r = E^{b}_r(r) + E^{p}_r(r, t), \,\,\ n^i = \frac{x^i}{\sqrt{x_ix^i}},
\end{equation}
where \(E^{b}_r(r)\) is a background radial electric field, found in \cite{Galushkina:2024knf}, and \(E^{p}_r(r, t)\) is a radial oscillating electric field of a perturbation mode. 
\begin{equation}
\label{E_field_back}
E^{b}_r(r)= - \frac{2 M^2 \gamma}{\Tilde{\alpha}} \times \frac{\vec{x}}{\sqrt{x_ix^i}}u(r)v(r)
\end{equation}
\begin{multline}
\label{E_field_pert}
E^{p}_r(r, t) = - \frac{2 M^2 \gamma}{\Tilde{\alpha}} \biggl[ u(r) (\phi_1(r) + \phi_2(r)) + \\
v(r) (\chi_1(r) + \chi_2(r))\biggr] cos(\lambda Mt)   
\end{multline}

Both background and perturbation electric fields decrease exponentially at spatial infinity, which is quite unusual for massless fields. In Fig.\ref{mode_electr} one can see profiles of the radial electric field, both for the background and perturbation mode, at \(\mathrm{w} = 0.9935\). 
Soliton consists of two shells, the field in which is directed oppositely. Non-topological solitons
with shells were considered earlier in \cite{Arodz:2008nm,Heeck:2021bce}.
Although locally \(j_{T} \neq 0\), the total electric charge of a solution is zero.

\section{Outlook}

In this paper, we consider a model describing interaction of neutral vector bosons with electromagnetic field via magnetic dipole moment. A special choice of interaction term in Lagrangian (\ref{L}) allows us to unambiguously link $F^{\mu \nu}$ and $W^{\mu \nu}$ at classical level for field configurations with trivial boundary conditions at spatial infinity, see Eq.(\ref{E_to_W2}). Because of this relation one can easily obtain localized bound state of electromagnetic field on vector soliton. In the NR limit, our consideration does not depend on the details of UV-completion. Possible scenarios of UV-completion were proposed in \cite{Loginov:2015rya,Herdeiro:2023lze}.

The studied mechanism of electromagnetic field localization can be naturally generalized to Proca stars. In this case, the stability of vector solitons is ensured by the gravitational attraction. Interestingly, recent progress in studying Proca stars \cite{Herdeiro:2023wqf} 
have drawn attention to scenarios of the non-spherical ground state.  
Being aware of this information it should be noted that a complete numerical study of non-spherical vector solitons will bring valuable knowledge about a variety of mechanisms of electromagnetic field localization. 

For Proca stars/balls, the search for quasinormal modes is of major interest \cite{Konoplya:2011qq, Ching:1998mxl}. As can be seen in Fig.\ref{lambda_w}, spherically symmetric perturbations serve as vibrational modes with discontinuous spectrum. This finding raise a question of following sort: what states lie in the region of delocalization? The authors suggest that seeking for quasinormal modes in this region might be promising in a future research.

Being not limited by phenomenology of astrophysical compact objects one may recall that theory (\ref{L}) is an effective field theory and thus could be useful for describing a system of non-relativistic ultra-cold atoms \cite{Lahaye:2009qqr}. In light of the results we present in this paper, the question arises about the possibility of experimental observation of electromagnetic field localization in a condensed matter.

\section{Acknowledgements}

The authors are indebted to Dmitry Levkov, Andrei Shkerin and Dmitry Galushkin for helpful discussions. Y.S.
would like to thank the Hanse-Wissenschaftskolleg Delmenhorst for support and hospitality. 
Numerical studies of this work were supported by the grant RSF 22-12-00215.

\appendix

\section{Supplementary material for Sec.\ref{Sec.3}}\label{app2}

In this appendix, we provide an explicit form of mathematical notations proposed in Eqs.(\ref{constr_perturb})
\begin{equation}
\label{constr_perturb_def}
    \begin{aligned}
& A_1 = (1 + (u^2 - v^2)) \chi_1  + u^2 (\chi_1 + \chi_2) - uv(\phi_1 + \phi_2) + \\
&\qquad + \kappa(- (u^2 - v^2)\chi_2 + 2 u^2 (\chi_1 + \chi_2) + 2 u v \phi_1) \;, \\
& A_2 = (1 + (u^2 - v^2)) \chi_2  + u^2 (\chi_1 + \chi_2) - uv(\phi_1 + \phi_2) + \\
&\qquad + \kappa(- (u^2 - v^2)\chi_1 + 2 u^2 (\chi_1 + \chi_2) + 2 u v \phi_2) \;, \\
& B_1 = (1 + (u^2 - v^2)) \phi_1  + uv (\chi_1 + \chi_2) - v^2(\phi_1 + \phi_2) - \\
&\qquad - \kappa((u^2 - v^2)\phi_2 + 2 v^2 (\phi_1 + \phi_2) + 2 u v \chi_1) \;, \\
& B_2 = (1 + (u^2 - v^2)) \phi_2  + uv (\chi_1 + \chi_2) - v^2(\phi_1 + \phi_2) - \\
&\qquad - \kappa((u^2 - v^2)\phi_1 + 2 v^2 (\phi_1 + \phi_2) + 2 u v \chi_2) \;, \\ 
& C_1 = 1 + u^2 - 2 (1 + \kappa) v^2 \;, \\ 
& C_2 = - (v^2 + \kappa(u^2 + v^2)) \;, \\ 
& D_1 =  2 (u u' - v v') \phi_1   - 2 vv' (\phi_2 + \phi_1)  + (vu' + uv') \times\\
&\qquad \times(\chi_2 + \chi_1) + u  v (\chi_2' + \chi_1')  - 2 \kappa (vv' \phi_1  + uu' \phi_2 + \\
&\qquad + (vu' + uv') \chi_1  + uv \chi_1' + vv' (\phi_2 + \phi_1)) \;, \\
& D_2 =  2 (u u' - v v') \phi_2  - 2 vv' (\phi_2 + \phi_1)  + (vu' + uv')\times\\
&\qquad \times(\chi_2 + \chi_1) + u  v (\chi_2' + \chi_1')  - 2 \kappa (vv' \phi_2  + uu' \phi_1 + \\
&\qquad + (vu' + uv') \chi_2  + uv \chi_2' + vv' (\phi_2 + \phi_1)). 
    \end{aligned}
\end{equation}

\section{Numerical methods}\label{app1}

In this appendix, we briefly discuss numerical methods of our study. We perform numerical integration on discretized lattice of size $L$ applying standard fourth-order Runge-Kutta scheme. For calculations near \(r = 0\), we use fields asymptotic coefficients at \(r \rightarrow 0\).

In order to find oscillating modes we apply two different methods. First, to analyze the spectrum in general, we use the modification of procedure, described in \cite{Levkov:2017paj}. We introduce a basis of two vector functions $\psi^{i}(r)=(\chi_{1}^{i}, \chi_{2}^{i}, \chi'^{i}_{1}, \chi'^{i}_{2},\phi_{1}^{i}, \phi_{2}^{i}), i=\{A,B\}$ with following conditions at \(r = L\)
\begin{equation}
    \begin{cases}
        &\psi^{A}(L)=(1, 0, A_{+}, 0, B_{+}, 0) \\
        &\psi^{B}(L)=(0,1,0,A_{-}, 0, B_{-}), \\
    \end{cases}
\end{equation}
where
\begin{equation}
\label{coef}
    \begin{aligned}
& A_{\pm} = \frac{w \pm \lambda}{\sqrt{1 - (w \pm \lambda)^2}} \times \left[ 1 + \frac{1}{r\sqrt{1 - (w \pm \lambda)^2} }\right]\; \\
& B_{\pm} = - \sqrt{1 - (w \pm \lambda)^2} \times \left[ 1 + \frac{1}{r\sqrt{1 - (w \pm \lambda)^2}  },\right]\; \\
    \end{aligned}
\end{equation}

These conditions allow for integrating equations of motion for linear perturbations basis functions from the right end of lattice to the center of soliton. For spherically symmetric perturbations the conditions in $r=0$ are
\begin{equation}
    \begin{cases}
        &\phi_{1}(0)=0;\\
        &\phi_{2}(0)=0.
    \end{cases}
\end{equation}
These conditions immediately lead to  \(\chi'_1, \chi'_2 = 0\) in the hedgehog background, see Eqs.(\ref{ans_eq_perturb_2}). 

Thus, we introduce a matrix $||D_{p}||$ that satisfies relation

\begin{equation}\label{matrix_D2}
\renewcommand\arraystretch{1.5}
    \begin{pmatrix}
\phi_{1}^{A}(0) & \phi_{1}^{B}(0) \\
\phi_{2}^{A}(0) & \phi_{2}^{B}(0)
\end{pmatrix} \begin{pmatrix}
    c_{A}\\
c_{B}\\
\end{pmatrix}=||D_{p}||\begin{pmatrix}
    c_{A}\\
c_{B}\\
\end{pmatrix}=0.
\end{equation}

\begin{figure}[h]
\includegraphics[width=1.0 \linewidth, height = 5.5 cm]{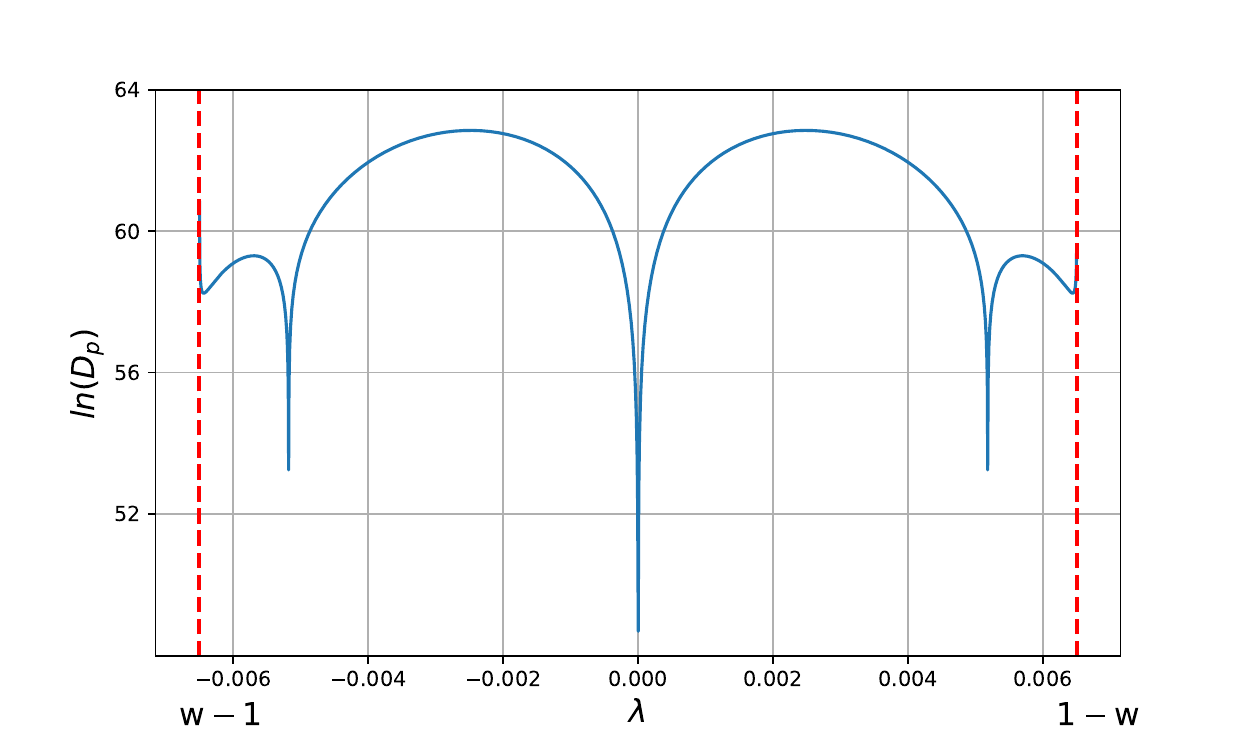}
\caption{Plotting $\ln{\det{||D_{p}||}}$ as a function of $\lambda$, the cusps mark the values of \(\lambda\) for localized states (\(\lambda_{loc} = \pm 0.00518\)).}
\label{scan}
\end{figure}

This algebraic system is solvable if $\det{||D_{p}||}=0$. This feature makes it possible to find parameter $\lambda$ of oscillating mode by plotting $\ln{\det{||D_{p}||}}$ as a function of $\lambda$, see Fig.\ref{scan}. Using this method, we obtain values of $\lambda$ accurate to the third significant digit.

Then, to check the results and draw the profiles of perturbations, we use a specific method which includes shooting. It is faster and more precise than scanning in the whole \(\lambda\) region, but depends on the analysis of spectrum made with the previous method. Once again, we introduce two basis vectors $\psi^{i}(r)=(\chi_{1}^{i}, \chi_{2}^{i}, \chi'^{i}_{1}, \chi'^{i}_{2},\phi_{1}^{i}, \phi_{2}^{i}), i=\{A,B\}$, but now they have initial conditions at \(r = 0\)
\begin{equation}
    \begin{cases}
        &\psi^{A}(0)=(1, 0, 0, 0, 0, 0) \\
        &\psi^{B}(0)=(0, 1, 0, 0, 0, 0), \\
    \end{cases}
\end{equation}
To avoid singularities in the numerical scheme, these functions can be calculated near zero point using asymptotic polynomial solutions at \(r\rightarrow0\).

At spatial infinity, each of these basis functions contains increasing exponents for two independent radial vector fields (\(\chi_1, \phi_1\) and \(\chi_2, \phi_2\)). For \(\lambda = \lambda_{loc}\) we can construct a function \(c_A \psi^{A} + c_B \psi^{B}\), which is a localized solution, so that increasing exponents compensate by the choice of coefficients. For \(\lambda \neq \lambda_{loc}\) we can compensate only one exponent, and the value of the second exponent at r = L can be used as a shooting criterion. Using this method, we find values of \(\lambda \) accurate to the 4th significant digit, check the results of previous method and plot resulting modes (see Fig.\ref{mode}).

\bibliographystyle{elsarticle-num} 
\bibliography{biblio}

\end{document}